\documentstyle[12pt]{article}
\def\fr{\frac}
\newcommand{\bq}{\begin{equation}}
\newcommand{\eq}{\end{equation}}
\begin{document}
\bigskip\bigskip
\bigskip
{\raggedleft {\makebox[2.7cm][1]{\it May. 1992}}\\}
\bigskip
\begin{center}

{\large\bf
Quantum  Liouville Theory On The Riemann Sphere With $n>3$ Punctures\\}

\bigskip\bigskip
Jian-Min Shen\\
\bigskip
{\small International Centre for Theoretical Physics, P. O. Box 586, 34100
Trieste, Italy.\\}

{\small and\\}

{\small Zhejiang Institute of Modern Physics, Zhejiang University, Hangzhou
310027, China. \footnote{Permanent Address.}}\\

\bigskip

Zheng-Mao Sheng\par

\bigskip

{\small Zhejiang Institute of Modern Physics, Zhejiang University, Hangzhou
310027, China.}\\

{\small and\\}

{\small Department of Physics, Hanzhou University, Hanzhou, China.$^1$\\}

\bigskip

Zhong-Hua Wang\\

\bigskip

{\small Institute of Theoretical Physics, Academia Sinica, Beijing 100080,
China.$^{1}$}\\

{\small and\\}

{\small SISSA, Strada Costiera 11, 34014 Trieste, Italy
\footnote{Mailing Address.}}\\

\bigskip\bigskip

\end{center}
\vskip 2pt

\medskip
\centerline{\bf ABSTRACT}
\bigskip\bigskip
\begin{center}
\begin{minipage}{120mm}
{\it
{}~~We have studied the quantum Liouville theory on the Riemann sphere
with n>3 punctures. While considering the theory on the Riemann surfaces
with n=4 punctures, the quantum theory near an
arbitrary but fixed puncture can be obtained via canonical quantization and
an extra symmetry is explored. While considering more than four
distinguished punctures, we have found the exchange relations of the
monodromy parameters from which we can get a reasonable quantum theory.}\\

\end{minipage}
\end{center}
\vskip 2pt
\bigskip\bigskip
{\bf PACS:02.20.+b}
\bigskip\bigskip\bigskip\bigskip
\vskip 2pt
\eject

\bigskip

{\raggedright \bf 1. Introduction\\}
\bigskip

  The Liouville theory has attracted much attention for a long time.
The early interests in it rests mainly on the uniformization theory of two
dimensional Riemann sphere and a lot of mathematicians such as Klein,
Poincare, Koebe [1] etc. have
done a lot of excellent works in this domain. The recent interests in it
revived mainly with the Polyakov string [2] and two dimensional quantum
gravity[3] where the Liouville action plays the role of Weyl anomaly. It is
inevitable for us to study the classical and quantum Liouville theory in order
to study the noncritical string theory.

  Despite the nice work for the Liouville theory4-5], there still exist a lot
of
mysterious aspects in both the classical and quantum Liouville theory.
In [6], we have mainly studied the classical
Liouville theory on the Riemann sphere with $n>3$ punctures and
interesting results have been obtained there.

  In this paper, we mainly study the quantum Liouville theory on the Riemann
sphere with $n>3$ punctures. Starting from the Poisson bracket relations, we
express these two chiral components which are two linearly independent
solutions
of the the uniformization equation in terms of the monodromy parameters and the
fields which depend only on the space-time coordinates.
Starting from the Poisson relations of the monodromy parameters and
the free field, we directly get the quantum Liouville theory in the
neighborhood
of a puncture via canonical quantization. In this case, we find
there exists an extra symmetry which leads to a certain arbitrariness for
the matrices dominating the classical and quantum exchange algebra relations.
If we restrict them to satisfy the classical and quantum Yang-Baxter equations
respectively, we find a family of solutions to both these equations.

  When we consider the theory on the riemann sphere with arbitrary number of
punctures, we will meet some monodromy parameters except the fields
depending only on the space time. We wish to get a quantum theory including
that
of the monodromy parameters and that of the space-time dependent fields.
To do this, we will adopt a postulation of the quantum exchange
properties of the monodromy parameters as well as the space-time dependent
fields. Starting from this postulation, we can get the quantum exchange algebra
relations of the chiral components from which we can see whether our
postulation
is reasonable or not since the quantum Liouville theory has the $SL(2, R)$
quantum group symmetry.

  This paper is organized as follows: In section 2, we introduce concisely,
for the convenience of notations, some background knowledge about the
uniformization theory of the Riemann sphere with $n>3$ punctures and some
main results of [6]. In section 3, we study the quantum Liouville theory,
via canonical quantization,
near an arbitrary but fixed puncture on the Riemann sphere. In section 4,
we study the quantum Liouville theory considering all the monodromy
parameters of $n>3$ punctures. Finally, we will give some concluding remarks.

\bigskip

{\raggedright \bf 2. Introduction to the Liouville theory on the Riemann
sphere with $n>3$ punctures.\\}

\bigskip

In this section, we will introduce some background knowledge about the
classical
Liouville theory and some main results of [7] for the convenience of
notation in the context.

  The Riemann sphere $X$ with $n$ punctures whose coordinates  are $z_1,
\cdots, z_{n-3}, 0, 1, and\\ \infty$ without loss of generality can be
realized as the quotient space $H/\Gamma$ where $H$
is the upper half plane and $\Gamma\subset PSL(2,~R)$ is the
Fuchsian group. That is to say, there exists a covering $J:~H \to X$ with
$J(\gamma \omega)=J(\omega)$ to arbitrary $\gamma\in \Gamma, ~\omega \in H$.

 The uniformization problem of the Riemann sphere is connected with the
differential equation of $J^{-1}(\omega)$. It has been shown [1] that $J^{-1}$
satisfy
\begin{equation}
Q_{X}=\fr{\partial_{\omega}^{3} J^{-1}}{\partial_{\omega} J^{-1}}
-\fr{3}{2}\Bigl (\fr{\partial_{\omega}^{2} J^{-1}}{\partial_{\omega} J^{-1}}
\Bigr)^{2}
=\sum_{i=1}^{n-1}\Bigl( \fr{1}{2(\omega-\omega_{i})^{2}}
+\fr{c_{i}}{\omega-\omega_{i}}\Bigr)
\end{equation}
where $c_{i}$ are the accessory parameters and satisfies
\begin{equation}
\sum_{i=1}^{n-1} c_{i}=0,~~~\sum_{i=1}^{n-1}c_{i}\omega_{i}=1-\fr{n}{2},
{}~~\sum_{i=1}^{n-1}\omega_{i}(1+c_{i}\omega_{i})=c_{n}
\end{equation}
which lead to the asymptotic behavior of $Q_{X}$
$$
Q_{X}=\fr{1}{2z^{2}} +\fr{c_{n}}{z^{3}}, ~~~~z\to \infty
$$

The projection monodromy group here is just the Fuchsian uniformazition group.
The elements of this group are parabolic if we only consider the punctured
Riemann surfaces.
We can choose in this group a standard system of parabolic
generators $M_{1}, \cdots, M_{n}$
satisfying the single relation $M_{1}\cdots M_{n}=1$, $Tr M_{\lambda}
=2$ and $M_{\lambda}$ has one fixed point $z_{\lambda}$, the matrices
satisfying these conditions can be represented as
\begin{equation}
M_{\lambda}=\left(\matrix{
1+\alpha_{\lambda}z_{\lambda} & -\alpha_{\lambda}z_{\lambda}^{2}\cr
\alpha_{\lambda} & 1-\alpha_{\lambda}z_{\lambda}\cr}\right)
\eq
and
\bq
M_{n}=\left(\matrix{
1 & \alpha_{n}\cr
0 &  1\cr}\right)
\eq
where we take $z_{n}=\infty$.

   Equation (1) is related to a second order linear differential equation
\begin{equation}
\fr{d^{2} \eta}{d z^{2}}+\fr{1}{2}Q_{X}(z)\eta=0
\end{equation}
by that $J^{-1}$ can be represented as the quotient of the two linearly
independent solutions of (5).

The poisson bracket relations between these two solutions of equation (5)
is dominated by
\begin{equation}
\{\eta_{i}(z), \eta_{j}(z')\}=
S_{ij}^{kl}\eta_{k}(z)\eta_{l}(z')
\end{equation}
where $i, j, k, l=1, 2$ and
\begin{equation}
S_{ij}^{kl}=-\fr{1}{16}[r_{+}\theta(|z|-|z'|)
+r_{-}\theta(|z'|-|z|)]_{ij}^{kl}
\end{equation}
and $\theta(z)$ is the step function. The $4\times 4$ matrices
$r_{\pm}$, called the classical $r$ matrices, are the solutions of the
classical
Yang-Baxter equation:
$$
[r_{12}, r_{13}]+[r_{12}, r_{23}]+[r_{13}, r_{23}]=0,
$$
and can be expressed by the generators of the Lie algebra $sl(2, R)$:
\begin{equation}
\begin{array}{ll}
r_{\pm}=\pm H\otimes H \pm 4E_{\pm}\otimes
E_{\mp}\\[4mm]
[H, E_{\pm}]= \pm 2E_{\pm},~~~~
[E_{+}, E_{-}]=H
\end{array}
\end{equation}

 When we circle around the $\lambda$-th puncture, $\eta_{1},~~\eta_{2}$
transform according to
\begin{equation}
\left(\matrix {\eta_{1}^{\lambda}\cr \eta_{2}^{\lambda}\cr}\right)
=\Bigl( M_{\lambda} \Bigr)\left(\matrix {\eta_{1}\cr \eta_{2}\cr}\right)
\end{equation}
Because the Liouville field $\phi$ is related to $\eta_{i}$ by
\begin{equation}
e^{-\fr{1}{2}\phi}=\fr{Im(\eta_{1}\bar{\eta_{2}})}{|\eta'_{1}\eta_{2}-
\eta_{1}\eta'_{2}|}
\end{equation}
so we can find that $\phi(P)$ is periodic for a closed path
$\Gamma_{\lambda}$ around any puncture on $z$ when the analytic
continuation of a pair of solutions $\eta_{1}(P)$ and $\eta_{2}(P)$  along
$\Gamma_{\lambda}$ results in  a pair of new solutions
$\eta_{1}^{\lambda}(P)$ and $\eta_{2}^{\lambda}(P)$.

That $\phi(z)$ is single valued  around puncture on $\Omega$ enables us to
make the following proposition:
\bq
\{\phi(P), \pi(P')\}=\{\phi(P+\Gamma_{\lambda}), \pi(P'+\Gamma_{\lambda})\}
=\triangle(P-P')
\eq
where $P, P'\in C_{\tau}$ and $C_{\tau}$ is the level curve, $\triangle(P-P')$
is the delta function on $C_{\tau}$. We define $C_{\tau}$ by
\begin{equation}
C_{\tau}=\{Q| Re\int_{Q_{0}}^{Q} \omega=\tau. \}
\eq
where $e^{\phi(z)}=\omega(z)\bar{\omega}(\bar{z})$ near the puncture.

  From (6) and (11), we can assume
\begin{equation}
\{\eta_{i}^{\lambda}(P), \otimes \eta_{j}^{\lambda}(P')\}
=S_{ij}^{kl}\eta_{k}^{\lambda}(P)\eta_{l}^{\lambda}(P')
\eq
from which the following Poisson brackets can be uniquely determined
\begin{equation}
\begin{array}{ll}
\{\eta_{1}, \alpha_{\lambda}\}=-\fr{1}{4}\alpha_{\lambda}z_{\lambda}\eta_{2},
{}~~~~
\{\eta_{1}, z_{\lambda}\}=-\fr{1}{8}z_{\lambda}^{2}\eta_{2},\\[4mm]
\{\eta_{2}, \alpha_{\lambda}\}=0, ~~~~~~~~~~~~
\{\eta_{2}, z_{\lambda}\}=-\fr{1}{8}\eta_{1}\\[4mm]
\{\alpha_{\lambda}, \alpha_{\rho}\}=0, ~~~~~~~~~~~~
\{z_{\lambda}, z_{\rho}\}=\fr{1}{8}(z_{\rho}^{2}-z_{\lambda}^{2})\\[4mm]
\{\alpha_{\lambda}, z_{\rho}\}=\fr{1}{4}\alpha_{\lambda}z_{\lambda}
\end{array}
\end{equation}
where $\lambda=1, ~\cdots, ~n-3~~and~n>3$ is the number of punctures.
Thus the elements of the monodromy matrices are dynamical variables in our
case.

\bigskip

{\raggedright \bf 3.  Liouville theory near one puncture on the Riemann
sphere with $n=4$ punctures.\\}

\bigskip

 Now we are ready to study the simplest case of our theory i.e. the
Liouville theory on the Riemann surfaces with $n=4$ punctures. Without loss
of generality, we may pick the coordinates of three of these punctures as
$0,~ 1,~ \infty$ and the another as $z_{\lambda}$. The monodromy matrices
corresponding to the $\lambda$-th puncture is as shown in equation (3).

 In this case, the fundamental relations in (14) can be rewritten in a more
compact form as
\begin{equation}
\begin{array}{ll}
\{q_{\lambda}, p_{\lambda}\}=1,~~~~~~
\{q_{\lambda}, q_{\lambda}\}=\{p_{\lambda}, p_{\lambda}\}=0,\\[4mm]
\{\eta_{1}, q_{\lambda}\}=-\fr{1}{2}e^{\fr{1}{4}(2p_{\lambda}-q_{\lambda})}
\eta_{2},~~~
\{\eta_{2}, q_{\lambda}\}=0,\\[4mm]
\{\eta_{1}, p_{\lambda}\}=0,~~~~~~~
\{\eta_{2}, p_{\lambda}\}=\fr{1}{4}e^{-\fr{1}{4}(2p_{\lambda}-q_{\lambda})}
\eta_{2},
\end{array}
\end{equation}
where
\bq
q_{\lambda}=2\ln \alpha_{\lambda},~~~~~
p_{\lambda}=\ln \alpha_{\lambda}+2\ln z_{\lambda}.
\eq

  From these relations, it is natural for us to define the Poisson bracket
relation in the monodromy parameter space in terms of $p_{\lambda}$ and
$q_{\lambda}$ as
\begin{equation}
\{A(q_{\lambda},~p_{\lambda}),~B(q_{\lambda},~p_{\lambda})\}_{P}
=\fr{\delta A}{\delta q_{\lambda}}\fr{\delta B}{\delta p_{\lambda}}
-\fr{\delta A}{\delta p_{\lambda}}\fr{\delta B}{\delta q_{\lambda}}
\end{equation}
where the index $P$ represents that the Poisson bracket is defined in the
monodromy parameter space.
With this explicit formulation, it is easy for us to get from (15) and (17)
these derivative equations satisfied by $\eta_{1}$ and $\eta_{2}$
\begin{equation}
\begin{array}{ll}
\fr{\delta \eta_{1}}{\delta p_{\lambda}}
=\fr{1}{2}e^{\fr{1}{4}(2p_{\lambda}-q_{\lambda})}\eta_{2},~~~~~~
\fr{\delta \eta_{2}}{\delta p_{\lambda}}=0,\\[4mm]
\fr{\delta \eta_{1}}{\delta q_{\lambda}}=0,~~~~~~~~~~~~~~~
\fr{\delta \eta_{2}}{\delta q_{\lambda}}
=\fr{1}{4}e^{-\fr{1}{4}(2p_{\lambda}-q_{\lambda})}\eta_{1},~~~
\end{array}
\end{equation}
By solving these equations, we get
\begin{equation}
\eta_{1}=e^{\fr{1}{2}p_{\lambda}}f(z),~~~~~
\eta_{2}=e^{\fr{1}{4}q_{\lambda}}f(z)
\end{equation}
where $f(z)$ is a function depending only on the space-time coordinates.

 It is worth noticing that $f(z)$ is a local solution of equation (5).
Since $J^{-1}$ can be represented as the quotient of two linearly independent
solutions of the Fuchsian equation (5), we find locally
\begin{equation}
\omega(z)=J^{-1}(z)=\fr{\eta_{1}}{\eta_{2}}= z_{i}
\end{equation}
according to equation (19).
It should be noticed that $J^{-1}$ in general is a multivalued analytic
function
on $X$. Suppose $(\eta'_{1},~\eta'_{2})$ is a pair of new solutions related
to the original pair of solution $(\eta_{1},~\eta_{2})$ by an arbitrary Mobius
transformation $\gamma\in \Gamma$. If $\gamma \not= M_{\lambda}$,
$J'^{-1}(z)=\fr{\eta'_{1}}{\eta'_{2}}$ has a branch different from that of
$J^{-1}(z)$ in equation (20). On the other hand, due to
$J(\gamma \omega)=J(\omega)$, $\omega\in H$, for any $\gamma\in \Gamma$, we
find
that $z_{i}$ corresponds to $\omega_{i}$ and $\gamma \omega_{i}$ on $H$ for any
$\gamma\in \Gamma$. On the other hand,
$J^{-1}$ is understood to depend on both the puncture's position and the
space-time coordinates. The expression (20) is a local one which holds in the
neighbourhood of the puncture $z_{\lambda}$.

  From (19), we may easily verify that $\eta_{1},~\eta_{2}$ are invariant
under the action of the monodromy group with generator (3). That is
$$
\begin{array}{ll}
\eta^{\lambda}_{1}=(1+\alpha_{\lambda}z_{\lambda})\eta_{1}
-\alpha_{\lambda}z_{\lambda}^{2}\eta_{2}=\eta_{1}\\[4mm]
\eta^{\lambda}_{2}=\alpha_{\lambda}\eta_{1}+
(1-\alpha_{\lambda}z_{\lambda})\eta_{2}=\eta_{2}
\end{array}
$$
Therefore $f(z)$ must be a local single valued function.

  Due to equations (6) and (17), the definition of the Poisson bracket
relation in (19) must be modified in the whole space $W$ spanned be
$p_{\lambda},~q_{\lambda}$ and
$f$. Since the Poisson bracket
between the monodromy parameters and $f$ equals zero, $W$ can be
regarded as the direct sum of two orthogonal subspaces spanned by
$p_{\lambda},~q_{\lambda}$ and
by $f$ respectively. The Poisson bracket relation between
$f$s at different positions can be obtained from
\bq
\{\eta_{i}(z),~ \eta_{i}(z')\}=-\fr{1}{16}\epsilon(|z|-|z'|)
\eta_{i}(z) \eta_{i}(z'),~~~i=1,~2
\eq
in (6) and (15) to be
\bq
\{f(z),~ f(z')\}=-\fr{1}{16}\epsilon(|z|-|z'|)
f(z) f(z')
\eq
If we denote $\varphi(z)=4\ln f(z)$, equation (22) turns to be
\bq
\{\varphi(z),~\varphi(z')\}=-\epsilon(|z|-|z'|).
\eq
So we can see that $f$ is similar to a vertex constructing from a free field
$\varphi$.

 Hence the Poisson bracket relation in $W$ in terms of the monodromy parameters
and $\varphi$ can be expressed as
\bq
\{A, ~B\}=\fr{\delta A}{\delta q_{\lambda}}\fr{\delta B}{\delta p_{\lambda}}
-\fr{\delta A}{\delta p_{\lambda}}\fr{\delta B}{\delta q_{\lambda}}
-\Bigl(\fr{\delta A}{\delta \varphi(z)}\fr{\delta B}{\delta\varphi(z')}
-\fr{\delta A}{\delta \varphi(z')}\fr{\delta B}{\delta \varphi(z)}\Bigr)
\epsilon(|z|-|z'|)
\end{equation}
where $A$ and $B$ are arbitrary functions smoothly depending on the monodromy
parameters and function $f$.

   From (19), we can find that there exists an extra symmetry in our theory,
that is
\begin{equation}
\eta_{1}(z)\eta_{2}(z')=\eta_{2}(z)\eta_{1}(z').
\end{equation}
This extra symmetry results from the monodromy symmetry around the puncture of
the theory and leads us to a modified classical exchange algebra relation.
\begin{equation}
\{\eta_{i}(z), \eta_{j}(z')\}=
\bar{S}_{ij}^{kl}\eta_{k}(z)\eta_{l}(z')
\end{equation}
where $i, j, k, l=1, 2$ and $\bar{S}$, in general, can be expressed as
\bq
\begin{array}{ll}
\bar{S}=-\fr{1}{16}
\left(\matrix{ 1 & 0 & 0 & 0 \cr
 0 & -1+l_{1} & 4-l_{1} & 0 \cr
0 & l_{2} & -1-l_{2} & 0\cr
0 & 0 & 0 & 1\cr}
\right)
{}~~~
\end{array}
\eq
where we have assumed $|z|>|z'|$ for simplicity without loss of generality and
$l_{1},~ l_{2}$ are some arbitrary constants.

  In this case, we will find that $\bar{S}$, in general, satisfies the
inhomogeneous Yang-Baxter equation:
\begin{equation}
[\bar{S}_{12},~\bar{S}_{13}]+[\bar{S}_{12},~ \bar{S}_{23}]+
[\bar{S}_{13},~\bar{S}_{23}]=\fr{1}{16^{2}} l_{2}(1-l_{1})
\epsilon_{ijk}A_{i}\otimes A_{j}\otimes A_{k}
\end{equation}
where $A_{1}=E_{+},~A_{2}=H,~A_{3}=E_{-}$ and $\epsilon_{ijk}=1$ for even
permutations of $123$ and $-1$ for odd permutations.

 In two special cases, i. e.

 (i): $l_{1}=1$, and $l_{2}$ is some arbitrary constant,

 (ii): $l_{2}=0$, and $l_{1}$ is some arbitrary constant,\\
we will get the ordinary classical Yang-Baxter equation. Hence, we have got
two classes of solutions of the classical Yang-Baxter equation by
substituting (i) and (ii) into the general expression (27):
\bq
\begin{array}{ll}
\bar{S}=-\fr{1}{16}
\left(\matrix{ 1 & 0 & 0 & 0 \cr
 0 & 0 & 3 & 0 \cr
0 & l_{2} & -1-l_{2} & 0\cr
0 & 0 & 0 & 1\cr}
\right)
{}~~~and ~~~~
\bar{S}=-\fr{1}{16}
\left(\matrix{ 1 & 0 & 0 & 0 \cr
 0 & -1+l_{1} & 4-l_{1} & 0 \cr
0 & 0 & -1 & 0\cr
0 & 0 & 0 & 1\cr}
\right)
\end{array}
\eq
When $l_{1}=l_{2}=0$, $\bar{S}$ in equation (27) is just the ordinary $SL(2,R)$
solution $r_{+}$ in (8) to the classical Yang-Baxter equation.

Now we are ready to study the quantum Liouville theory on the Riemann sphere
with four punctures. The method to the quantization is the usual canonical
quantization, i.e. we first regard all the classical functions such
as $\eta_{1}, ~\eta_{2}, ~p_{\lambda},~ q_{\lambda}, ~\varphi$ in (15), (23)
and
etc. as operators, then we replace the canonical Poisson brackets of
$q_{\lambda}$ and $p_{\lambda}$ in (15) and that of
$\varphi$ in (23) by commutators according to the quantization principle,
i.e. in the quantum case, we have these commutation relations:
\begin{equation}
\begin{array}{ll}
[q_{\lambda}, p_{\lambda}]=i\hbar,
{}~~~[p_{\lambda},~\varphi(z)]=[q_{\lambda},~\varphi(z)]=0
\\[4mm]
[\varphi(z),~\varphi(z')]=-i\hbar\epsilon(|z|-|z'|)~~~
\end{array}
\eq

  In the quantum case, $\eta_{1}$ and $\eta_{2}$ can be defined as
$$
\eta_{1}=:e^{\fr{1}{2}p_{\lambda}}::e^{\fr{1}{2}\varphi}:,~~~~~
\eta_{2}=:e^{\fr{1}{4}q_{\lambda}}::e^{\fr{1}{2}\varphi}:
$$
where $:~:$ denotes normal ordering. With this definition, we can find,
corresponding to the classical case, there exists in the quantum case an extra
symmetry:
\begin{equation}
\eta_{1}(z) \eta_{2}(z')
=e^{-\fr{1}{8} i\hbar}\eta_{2}(z) \eta_{1}(z')
\eq
which leads to these exchange algebra relation:
\bq
\eta_{i}(z)\eta_{j}(z')=R_{ij}^{mn}\eta_{m}(z')\eta_{n}(z)
\eq
where
\bq
R=e^{-\fr{i}{16}\hbar}
\left(\matrix{
1 & 0 & 0 & 0\cr
0 & k_{1}e^{\fr{i}{4}\hbar} & e^{-\fr{i}{8}\hbar}- k_{1}e^{\fr{i}{8}\hbar}
& 0\cr
0 & e^{\fr{i}{8}\hbar}-k_{2} & k_{2}e^{\fr{-i}{8}\hbar} & 0\cr
0 & 0 & 0 & 1 \cr
}\right)
\eq
and $k_{1}$, $k_{2}$ are some quantities to be determined.

  Before we set to find the equation satisfied by (33), we first show that
if $k_{1}=\fr{1-l_{1}}{16}i\hbar+ O(\hbar^{2})$ and
$k_{2}=\fr{1-l_{2}}{16}i \hbar+O(\hbar^{2})$ when $\hbar \to 0$, then we have
\begin{equation}
R=P_{12} I\otimes I++i\hbar \bar{S}+O(\hbar^{2})
\end{equation}
where $P_{12}$ is the permutation operator of $1$ and $2$ and $\bar{S}$ is
as shown in  equation (27). Hence we may say that (33) can be regarded as
the quantum counterpart of the $\bar{S}$ matrice in (27).

For general $k_{1}$ and $k_{2}$, $R$ satisfies the following inhomogeneous
quantum Yang-Baxter equation:
\begin{equation}
R_{12}R_{13}R_{23}-R_{23}R_{13}R_{12}=k_{1}k_{2}(\fr{1}{2}I\otimes(A+D)
+\fr{1}{2}H\otimes (A-D)+E_{-}\otimes B +E_{+}\otimes C)
\end{equation}
where
$$
\begin{array}{ll}
A=(a^{3}-\fr{k_{2}}{a})(E_{-}\otimes E_{+}-E_{+}\otimes E_{-})+\fr{1}{4}
(\fr{k_{2}}{a^{3}}-a)(I-H)\otimes (I+H)\\[4mm]
B=a(a^{2}k_{1}-1)H\otimes E_{+}+
\fr{1}{2a}(1-a^{2}k_{1}-a^{4}+k_{2})E_{+}\otimes I +
\fr{1}{2a}(1-a^{2}k_{1}+a^{4}-k_{2})E_{+}\otimes H\\[4mm]
C=(a-\fr{k_{2}}{a^{3}})H\otimes E_{-}
-\fr{1}{2a}(1-a^{2}k_{1}-a^{4}+k_{2})E_{-}\otimes I
-\fr{1}{2a}(1-a^{2}k_{1}+a^{4}-k_{2})E_{-}\otimes H\\[4mm]
D=\fr{1}{4}(a^{3}k_{1}-\fr{k_{2}}{a^{3}})(I+H)\otimes (I-H)
+\fr{1}{a}(a^{2}k_{1}-1)(E_{-}\otimes E_{+}-E_{+}\otimes E_{-})
\end{array}
$$
and $a=e^{-\fr{i}{16}\hbar}$.

In these special cases

(i): $k_{1}=0$, for arbitrary $k_{2}$.

(ii): $k_{2}=0$, for arbitrary $k_{1}$.

(iii): $k_{2}=a^{4}$, $k_{1}=a^{-2}$.\\
 equation (37) reduces to the ordinary Quantum Yang-Baxter equation. So
three kinds of matrices
$$
R=e^{-\fr{i}{16}\hbar}
\left(\matrix{
1 & 0 & 0 & 0\cr
0 & 0 & e^{-\fr{i}{8}\hbar} & 0\cr
0 & e^{\fr{i}{8}\hbar}-k_{2} & k_{2}e^{\fr{-i}{8}\hbar} & 0\cr
0 & 0 & 0 & 1 \cr
}\right)
 ~~and~~
R=e^{-\fr{i}{16}\hbar}
\left(\matrix{
1 & 0 & 0 & 0\cr
0 & k_{1}e^{\fr{i}{4}\hbar} & e^{-\fr{i}{8}\hbar}- k_{1}e^{\fr{i}{8}\hbar}
& 0\cr
0 & e^{\fr{i}{8}\hbar} & 0 & 0\cr
0 & 0 & 0 & 1 \cr
}\right)$$
in our case satisfy the quantum Yang-Baxter equation:
$$
R_{12}R_{13}R_{23}=R_{23}R_{13}R_{12}
$$

\bigskip

{\raggedright \bf 4. General Case.\\}

\bigskip

In this section, we consider the Liouville theory considering
arbitrary number of punctures.

Let us first reparametrize the monodromy parameters
$\alpha_{\lambda},~ z_{\lambda}$ as
$$
\alpha_{\lambda}^{\fr{1}{2}}=Q_{\lambda},~~~~~~~~~~
z_{\lambda}\alpha_{\lambda}^{\fr{1}{2}}=P_{\lambda}.
$$
Then the Poisson bracket relation (14) can be rewritten as
\bq
\begin{array}{ll}
\{Q_{\lambda},~P_{\rho}\}=\fr{1}{8}P_{\lambda}Q_{\rho},~~~~~~
\{Q_{\lambda},~Q_{\rho}\}=0\\[4mm]
\{\eta_{1},~ Q_{\lambda}\}=-\fr{1}{8}P_{\lambda}\eta_{2},
{}~~~~~~~~~~\{P_{\lambda},~P_{\rho}\}=0\\[4mm]
\{\eta_{1},~ P_{\lambda}\}=0,~~~~~~~~~~~~~~~~~\{\eta_{2},~ Q_{\lambda}\}=0
\\[4mm]
\{\eta_{2},~ P_{\lambda}\}=\fr{1}{8}Q_{\lambda}\eta_{1}
\end{array}
\end{equation}

Comparing with the notation of the last section in the neighbourhood of the
$\lambda$-th puncture, we have
$$
P_{\lambda}=e^{\fr{1}{2}p_{\lambda}},~~~~~~~~
Q_{\lambda}=e^{\fr{1}{4}q_{\lambda}}
$$

Generally, if we consider the Poisson bracket relation between the
functions $F=F(Q_{i}.P_{i})$ and $G=G(Q_{i}.P_{i})$, we can get
from equation the property of the Poisson bracket and (36),
\begin{equation}
\begin{array}{ll}
\{F,~G\}=
\fr{\partial F}{\partial Q_{i}}\fr{\partial G}{\partial Q_{j}}\{Q_{i},~Q_{j}\}+
\fr{\partial F}{\partial Q_{i}}\fr{\partial G}{\partial P_{j}}\{Q_{i},~P_{j}\}+
\fr{\partial F}{\partial P_{i}}\fr{\partial G}{\partial Q_{j}}\{P_{i},~Q_{j}\}+
\fr{\partial F}{\partial P_{i}}\fr{\partial G}{\partial P_{j}}\{P_{i},~P_{j}\}
\\[4mm]
=\fr{1}{8}[\bigtriangledown_{Q} F \cdot \bigtriangledown_{P} G-
\bigtriangledown_{P} F \cdot \bigtriangledown_{Q} G]
\end{array}
\end{equation}
where
$$
\bigtriangledown_{Q}=\sum_{i}P_{i}\fr{\partial}{\partial Q_{i}},~~~~~~~
\bigtriangledown_{Q}=\sum_{i}Q_{i}\fr{\partial}{\partial P_{i}}.
$$
Equation (37) may also be regarded as the defining relation for the poisson
bracket in the parameter space spanned by $P_{i},~Q_{i}$.

{}From equation (14) and the defining relation (37), we can easily get the
differential equations satisfied by $\eta_{1}$ and $\eta_{2}$ in this
parameter space:
\begin{equation}
\begin{array}{ll}
\sum_{i=1}^{n-3}Q_{i}\fr{\partial \eta_{1}}{\partial P_{i}}=\eta_{2},~~~~
\sum_{i=1}^{n-3}P_{i}\fr{\partial \eta_{2}}{\partial Q_{i}}=\eta_{1},\\[4mm]
\sum_{i=1}^{n-3}P_{i}\fr{\partial \eta_{1}}{\partial Q_{i}}=0,~~~~
\sum_{i=1}^{n-3}Q_{i}\fr{\partial \eta_{2}}{\partial P_{i}}=0
\end{array}
\end{equation}
The general solution to this equation can be found to be [7]
\begin{equation}
\eta_{1}=\sum_{i=1}^{n-3}f_{i}(z)P_{i},~~~~~~
 \eta_{2}=\sum_{i=1}^{n-3}f_{i}(z)Q_{i}
\end{equation}
where $f_{i}(z)$ depends only on the space-time coordinates and must
satisfies the uniformization equation (1).

If we consider two functions $F=F(P_{i}, Q_{i}, f_{i}, f_{i}')$ and
$G=G(P_{i}, Q_{i}, f_{i}, f_{i}')$ in which $f_i=f_i(z), ~f_i'=f_i(z')$, their
Poisson bracket can be found to be
\begin{equation}
\{F,~ G\}=\{F,~G\}_{(Q_{i},~P_{j})}\cdot \{Q_{i},~P_{j}\}+
\{F,~G\}_{(f_{i},~f_{j}')}\cdot \{f_{i},~f_{j}'\}
\end{equation}
where
$$
\begin{array}{ll}
\{F,~G\}_{(Q_{i},~P_{j})}=
\fr{\partial F}{\partial Q_{i}}\cdot\fr{\partial G}{\partial P_{j}}
-\fr{\partial F}{\partial P_{j}}\cdot\fr{\partial G}{\partial Q_{i}}\\[4mm]
\{F,~G\}_{(f_{i},~f_{j}')}=
\fr{\partial F}{\partial f_{i}}\cdot\fr{\partial G}{\partial f_{j}'}
-\fr{\partial F}{\partial f_{j}'}\cdot\fr{\partial G}{\partial f_{i}}
\end{array}
$$

 From (6) and (39),  we can
get the Poisson bracket relations between $f_{i}$ and $f_{i}'$ as
\begin{equation}
\{f_{i}, f_{j}'\}=\fr{1}{16}\epsilon(|z|-|z'|)(f_{i}f_{j}'-2f_{j}f_{i}')
\end{equation}

 When we consider the theory on the Riemann surface with more than four
punctures, one may notice that neither the parameters $P_i$, $Q_i$
nor the space-time dependent fields $f$, $f'$ in (39) can be considered as
free fields.

However, in the canonical quantization, when we try to get the quantum theory
corresponding to the classical theory, we have made an important assumption--
the principle of quantization. This assumption corresponding to the
canonical Poisson bracket relation between the canonical variables is the
bridge between the classical and quantum theory.

 What we face now is: we have not the canonical conjugate variables at hands.
So the  first thing we should do is to find a principle of quantization
corresponding to our Poisson bracket relations. We assume that all the
functions such as $\eta_{1},~\eta_{2}, f(z)$ and all the monodromy
parameters such as $Q_{i}, ~P_{i}$ will be regarded as operators in the
quantum case. The expression of $\eta_{1},~\eta_{2}$ in terms of $f(z),~
Q_{i}, ~P_{i}$ is the same as its classical case in equation (39).

Furthermore, we make the following principle of quantization:
\begin{equation}
\begin{array}{ll}
P_{k}Q_{l}=Q_{l}P_{k}+(1-A)P_{l}Q_{k}\\[4mm]
f_{l}f_{k}'=(2-B)f_{k}'f_{l}-2(1-B)f_{l}'f_{k}
\end{array}
\end{equation}
where $A$ and $B$ are some coefficients to be determined (See [7] for
details).

Furthermore, we can get from (39) and (42) these exchange relations
\begin{equation}
\begin{array}{ll}
\eta_{1}\eta_{1}'=B\eta_{1}'\eta_{1},~~~~~
\eta_{1}\eta_{2}'=(2-B)\eta_{2}'\eta_{1}+(B-2A+AB)\eta_{1}'\eta_{2},\\[4mm]
\eta_{2}\eta_{2}'=B\eta_{2}'\eta_{2},~~~~~
\eta_{2}\eta_{1}'=(2-B)A(2-A)\eta_{1}'\eta_{2}-(4-3B-2A+AB)\eta_{2}'\eta_{1}.
\end{array}
\end{equation}
They can be rewritten compactly as
\begin{equation}
\eta_{i}\eta_{j}'=R_{ij}^{mn}\eta_{m}\eta_{n}'
\end{equation}
where
\begin{equation}
R=\left(\matrix{
B & 0 & 0 & 0\cr
0 & 2-B & B-2A+AB & 0 \cr
0 & -4+3B+2A-AB & (2-B)A(2-A) & 0 \cr
0 & 0 & 0 & B\cr}\right)
\eq

If $A=\fr{4-3B}{2-B}$, we will get
\begin{equation}
R_{+}=\left(\matrix{
B & 0 & 0 & 0\cr
0 & 2-B & -4+4B & 0 \cr
0 & 0 & \fr{B(4-3B)}{2-B} & 0 \cr
0 & 0 & 0 & B\cr}\right).
\eq

If $A=\fr{B}{2-B}$, we will get
\begin{equation}
R_{-}=\left(\matrix{
B & 0 & 0 & 0\cr
0 & 2-B & 0 & 0 \cr
0 & -4+4B & \fr{B(4-3B)}{2-B} & 0 \cr
0 & 0 & 0 & B\cr}\right).
\eq

It is easy to show that both $R_{+}$ and $R_{-}$ satisfies the quantum
Yang-Baxter equation.

If we let $B=exp(\fr{i\hbar}{16})$, we find
\begin{equation}
R_{+}|_{\hbar\to 0}=I+\fr{i\hbar}{16}\left(\matrix{
1 & 0 & 0 & 0\cr
0 & -1 & -4 & 0 \cr
0 & 0 & -1 & 0 \cr
0 & 0 & 0 & 1\cr}\right)+o(\hbar^{2})=I+\fr{i\hbar}{16}r_{+}+o(\hbar^{2}).
\\[4mm]
\end{equation}
If we let $B=exp(-\fr{i\hbar}{16})$, we find
\begin{equation}
R_{-}|_{\hbar\to 0}=I-\fr{i\hbar}{16}\left(\matrix{
1 & 0 & 0 & 0\cr
0 & -1 & 0 & 0 \cr
0 & 4 & -1 & 0 \cr
0 & 0 & 0 & 1\cr}\right)+o(\hbar^{2})=I-\fr{i\hbar}{16}r_{-}+o(\hbar^{2}).
\eq

\bigskip
{\raggedright \bf 5. Concluding Remarks.\\}
\bigskip

At this stage, we have got the quantum Liouville theory on the Riemann
sphere with $n>3$ punctures. To make the problems clearer, let us
present the logic as follows: we are studying the Liouville theory on the
Riemann sphere with $n>3$ punctures, the chiral components which are two
independent solutions of the uniformization equation depend in general on
the space-time coordinates and the monodromy parameters, its classical
exchange algebra relations is dominated by the classical $r$-matrix and its
quantum exchange algebra relation, in principle, can be assumed to be
dominated by the quantum $R$-matrices. This is in fact the quantum
integrability condition. When we only consider the quantum theory in the
neighbour of an arbitrary but fixed puncture, we can find the canonical
variables in the monodromy parameter space and space dependent field space.
We can get the quantum theory from the classical one via canonical
quantization. When we consider more then four distinguished punctures,
we find that the usual canonical quantization procedure does not do. However,
 we want to get the quantum theory of
all the dynamical variables which include the monodromy parameters. For
this reason, we adopt the assumption of
the quantum exchange relations of the monodromy parameters and that of the
space-dependent fields, i.e. equation (42). Then we try to get the
quantum exchange algebra relations between the chiral components according
to this assumption. If this assumption can lead to the
quantum integrability condition, we can say that this assumption is
reasonable and hence we have got the quantum exchange relations between all
the dynamical variables.

\bigskip

\bigskip
\bigskip
{\raggedright \bf Acknowledgement\\}
\bigskip

  The authors are obliged to Prof. L. Bonora, H. Y. Guo, A. Verjovsky and
R. Wang for their stimulating discussions and encouragement in the course of
this work. J. M. Shen would like to thank
Prof. Abdus Salam, the International Atomic Energy Agency and UNESCO for
hospitality at the International Centre for Theoretical Physics, Trieste,
Italy. Z. H. Wang is obliged to Profs. D. Amati, L. Bonora and R. Iengo
for their hospitalities during his works in SISSA sponsored by the World
Laboratory.

\bigskip
{\raggedright \bf References\\}
\begin{description}
\bigskip
\item[1] H. Poincar\'e, J. Math. Pures. Appl. (5)4(1898) 137-230.
         P. Zograf and L. Takhtajan, Math. USSR sbornik, Vol. 60(1988),
         No. 1, 143-161. and references therein.
\item[2] A. M. Polyakov, Phys. Lett. B. 103(1981)207.
\item[3] A. M. Polyakov, Mod. Phys. Lett. A2 (1987)893; V.G. Knizhnik, A.M.
        Polyakov and A. B. Zamolodchikov, Mod. Phys. Lett. A3(1988)819;
       F. David, Mod. Phys. Lett. A3 (9188)1651; J. Distler and H. Kawai, Nucl.
       Phys. B321(1989)509.
\item[4]J. L. Gervais and A. Neveu, Nucl. Phys. B199(1982) 59-76;
B209(1982)125;
B238(1984)125. Babelon, Phys. Lett. B215(1988)523. J. L. Gervais, Comm.
Math. Phys. 130(1990)257, 138(1991)301.
\item[5] T. Curtright and C. Thorn. Phys. Rev. Lett. 48(1982)1309; E. Braaten,
T. Curtright, G. Ghandour and C. Thorn. Phys. Rev. Lett. 51(1983)19.
\item[6] J. M. Shen and Z. M. Sheng, ICTP Preprint, IC/91/376. To appear in
Phys. Lett. B. (1992).
\item[7] J. M. Shen, Z. M. Sheng and Zhong-Hua Wang ICTP Preprint, IC/92/23.

\end{description}

\end{document}